\documentclass[preprint,showpacs,preprintnumbers,amsmath,amssymb,nofootinbib]{revtex4}
\def \be {\begin{equation}}
\def \ee {\end{equation}}
\def \bea {\begin{eqnarray}}
\def \eea {\end{eqnarray}}

\usepackage[dvips]{graphicx,color}
\usepackage{dcolumn}
\usepackage{bm}
\usepackage{epsfig}

\begin{document}

\title{Non-Gaussian statistics, Maxwellian derivation and stellar polytropes}

\author{E. P. Bento$^{1}$}
\email{eliangela@dfte.ufrn.br}

\author{J. R. P. Silva$^{2}$}
\email{joseronaldo@uern.br}

\author{R. Silva$^{1,2}$}
\email{raimundosilva@dfte.ufrn.br}

\affiliation{$^{1}$Universidade Federal do Rio Grande do Norte, UFRN,
Departamento de F\'{\i}sica
C. P. 1641, Natal -- RN, CEP 59072--970, Brazil}

\affiliation{$^{2}$Universidade do Estado do Rio Grande do Norte, UERN,
Departamento de F\'{\i}sica, Mossor\'o -- RN, CEP 59610--210 , Brazil}

\date{\today}

\begin{abstract}
In this letter we discuss two aspects of non-Gaussian statistics. In the first, we show that Maxwell's first derivation of the stationary distribution function for a dilute gas can be extended in the context of Kaniadakis statistics. In the second, by investigating the stellar system, we study the Kaniadakis analytical relation between the entropic parameter $\kappa$ and the stellar polytrope index $n$. We compare also the Kaniadakis relation $n=n(\kappa)$ with $n=n(q)$ proposed in the Tsallis framework.

\end{abstract}

\maketitle

\section{Introduction}


Nonextensive statistical mechanics (NSM) \cite{tsallis1} and extensive
generalized power-law statistics \cite{k01,k02,k03,review,k10,k11} are based on the mathematical generalization of the Boltzmann exponential distribution, namely:
\begin{equation}
f\sim\exp(-Energy/Thermal\quad energy).
\end{equation}
Whereas in \cite{tsallis1}, the power-law distribution, $f\sim[1-(1-q)Energy/Thermal\quad energy]^{1/(1-q)}$, obeys the Tsallis distribution of NSM \cite{tsallis1}, the so-called
$\kappa$-statistics \cite{k01,k02,k03,review,k10,k11} provides a power-law
distribution function and the $\kappa$-entropy which emerges in the
context of the special
relativity and in the {\it kinetic interaction principle}. Formally, the $\kappa$-framework is defined by considering the expressions
\begin{equation}\label{expk}
\exp_{\kappa}(f)= (\sqrt{1+{\kappa}^2f^2} + {\kappa}f)^{1/{\kappa}},
\end{equation}
\begin{equation}\label{expk1}
\ln_{\kappa}(f)= {{f^{\kappa}-f^{-\kappa}\over 2\kappa}},
\end{equation}
where the $\kappa$-entropy associated with $\kappa$-statistics is
given by
\cite{k01,k02,k03,review,k10,k11}
\begin{equation}\label{first}
S_\kappa = - \int d^3 p f\ln_\kappa f =  - \langle{\ln_\kappa
(f)\rangle}.
\end{equation}
The expressions above reduce to the standard results in the limit $\kappa=0$.

$\kappa$-statistics lead to a {\bf generalized framework} with interesting mathematical properties \cite{pla01,phya02,physicaa02}, as well as a connection with the
generalized Smoluchowski equation \cite{chavanis04} and the relativistic nuclear equation of state for nuclear matter
\cite{pereira09}. It was also shown
that it is possible to obtain
a consistent form for the entropy which is linked with a two-parameter
deformation of the logarithm
function and generalizes the Tsallis, Abe and Kaniadakis logarithm
behaviours \cite{k4}. {\bf Some physical systems are well approximated by a distribution that maximizes Kaniadakis' entropy}, namely,
cosmic ray flux, rain events in meteorology \cite{k02}, quark-gluon
plasma \cite{Teweldeberhan03}, kinetic models
describing a gas of interacting atoms and photons \cite{rossani04}, fracture
propagation phenomena \cite{Cravero04}, and construct financial
models \cite{Rajaonarison05}. In the theoretical front, some studies on the
canonical quantization of a classical system have also been investigated
\cite{scarfone05}, as well as the $H$-theorem in relativistic and non-relativistic
domain \cite{silva06,silva2006}.

In the astrophysical domain, the first application has been
the simulation in relativistic plasmas. In this regard, the power-law
energy distribution provides
a strong argument in favour of the Kaniadakis statistics \cite{lapenta07}. Additionally, the viability of non-Gaussian statistics has been investigated from a stellar astrophysics viewpoint: the distributions of projected rotational velocity
measurements of stars in the Pleiades open cluster \cite{carvalho08}, in the main sequence field stars \cite{carvalho09}, {\bf and in the estimation of the mean angle of inclination of the rotational axes of the stars in the Orion Nebula Cloud \cite{soares11}}, as well as the strong dependence between the stellar-cluster ages and the power-law distributions \cite{epl10}.

The aim of this letter is twofold. First, to show that the Maxwell's first derivation of the stationary distribution function for a
dilute gas can be extended in the context of Kaniadakis statistics; Second, considering the principle of maximum entropy for a stellar self-gravitating system, to investigate an analytical relation between entropic parameter $\kappa$ and stellar polytrope index $n$. It is also show that the function $n=n(\kappa)$ has a similar behaviour to Tsallis expression $n=n(q)$ \cite{plastino93,plastino05}.

This letter is organized as follows. In Section 2, we show the
correspondence between the $\kappa$-statistics introduced by Kaniadakis and the velocity
distribution for a Maxwellian gas, by assuming a non-Gaussian generalization of the separability hyphothesis originally proposed by Maxwell \cite{maxwell60}. In Section 3, we discuss a connection between Kaniadakis statistics and the polytropic index in the context of the self-gravitating system and we compare our results with ones studied in Ref. \cite{plastino93,plastino05}. Finally, Section 4 is devoted to conclusions and discussion.

\section{Non-Gaussian Maxwellian distribution function}

In this section, in order to introduce the generalization of the Maxwell distribution in the context of Kaniadakis framework. Let us consider a spatially homogeneous gas, assumed to be in equilibrium (or in stationary state) at temperature
$T$, in such a way that $F(v)d^3v$ is the number of particles with velocity $v$ in the volume element
$d^3v$ around $v$. In Maxwell's derivation, the 3-dimensional distribution is factorized and depends only on the magnitude of the velocity \cite{maxwell60,sommerfeld93}
\begin{equation}\label{maxwellian}
F(v)d^3v=f(v_x)f(v_y)f(v_z)dv_xdv_ydv_z\;\;,
\end{equation}
where $v=\sqrt{v_x^2+v_y^2+v_z^2}$ and $F(v)$ is the standard Maxwellian
distribution function, given by
\begin{equation}\label{1}
F(v)=A\exp( -{m v^2 / k_BT} ),
\end{equation}
where $A=\left( {m\over 2\pi k_B T} \right)^{3/2}$ is the normalization constant.

In reality, in the $\kappa$-statistics context described
by (\ref{first}), the starting basic hypothesis (\ref{maxwellian}), which takes into account the isotropy of all velocity
directions, must somewhat be modified. From a statistical viewpoint, Maxwell's ansatz assumes that the three components of the velocity are {\bf statistically independent}.
However, this property does not hold in the systems endowed with long range interactions, or statistically dependent, where Kaniadakis distribution
has been observed \cite{carvalho08,carvalho09}. Notice that the Maxwell ansatz is equivalent to expressing $\ln F$
as the sum of the logarithms of the one dimensional distribution functions associated with
each velocity component. A simple and natural way to generalize this procedure within the
Kaniadakis framework, {\bf would be to introduce  statistical dependence between the velocity components},
e.g., to replace {\bf the usual product between $f(v_x)$, $f(v_y)$ and $f(v_z)$ by a $\kappa$-exponential of the sum of $\ln_{\kappa}$ of the $f(v_i)$, $i=x,y,z$. From a physical viewpoint, the statistical dependence allows introducing a distribution that has a better fit than the Maxwellian in the statistical description of some
physical systems (See, e.g., Refs. \cite{pereira09,lapenta07,carvalho08,carvalho09,epl10})\footnote{\bf Obviously this is not a unique generalization. For example, in the Tsallis framework it is possible to introduce statistical dependence between velocity components considering the $q$-generalization of the Maxwell ansatz (See, e.g, Ref \cite{silva98})}}. Therefore, in order to recover
the ordinary logarithmic ansatz as a particular limiting case, it is convenient to express
the power generalization in terms of the function $\ln_\kappa$ defined by Eq. (\ref{expk1}), which is a combination of a power
function plus appropriate constants. Mathematically, the consistent $\kappa$-generalization of (\ref{maxwellian}) is given by
\begin{eqnarray}\label{kdistribution} \nonumber
F\left(\sqrt{v_x^2+v_y^2+v_z^2}\right)d^3v &=& \exp_\kappa[\ln_\kappa {f(v_x)}+\ln_\kappa {f(v_y)}+\\
&&\hspace{0.2truecm}\ln_\kappa {f(v_z)}]dv_xdv_ydv_z\;,
\end{eqnarray}
where the $\kappa$-exponential and $\kappa$-logarithm are given by
identities (\ref{expk}) and (\ref{expk1}).
In particular, in the limit $\kappa=0$ the standard expressionr
(\ref{maxwellian}) is recovered. Note also that
$\ln_\kappa[\exp_\kappa(f)]=\exp_\kappa[\ln_\kappa(f)]=f$, and
$\frac{d{\ln_\kappa(f)}}{dx}=\frac{f^\kappa+f^{-\kappa}}{2f}\frac{df}{dx}$
are satisfied.
The logarithmic derivative of (\ref{kdistribution}) with respect to $v_i$ is
\begin{eqnarray}\label{kderivate} \nonumber
\frac{\partial }{\partial v_i}\ln_\kappa
F\left(\sqrt{v^2_x+v^2_y+v^2_z}\right)&=&\frac{\partial }{\partial v_i}\ln_\kappa
\{\exp_\kappa [\ln_\kappa f(v_x)+\\
&& \ln_\kappa f(v_y)+\ln_\kappa f(v_z)]\},
\end{eqnarray}
with $i=\emph{x,y,z}$.

Using the above mentioned properties we can write\footnote{\bf It is worth mentioning that repeated index does not mean a summation over the index.}
\begin{equation}
\frac{F^\kappa+F^{-\kappa}}{2F}\frac{\partial}{\partial
v_i}F\left(\sqrt{v^2_x+v^2_y+v^2_z}\right)=\frac{\partial }{\partial v_i}[\ln_\kappa
f(v_i)]\;\;.
\end{equation}

Equivalently,
\begin{equation}\label{quieq}
\frac{F^\kappa+F^{-\kappa}}{2F}\frac{F'(\chi)}{\chi}=\frac{1}{v_i}\frac{\partial}{\partial
v_i}\ln_\kappa f(v_i)\;\;,
\end{equation}
where $\chi=\sqrt{v^2_x+v^2_y+v^2_z}$ {\bf and a prime represents the total derivative}. Now, defining
\begin{equation}\label{phieq}
\Phi(\chi)=\frac{F^\kappa+F^{-\kappa}}{2F}\frac{F'(\chi)}{\chi}\;\;,
\end{equation}
{\bf and considering the replacement of the partial derivative by an ordinary derivative in Eq. (\ref{quieq}) due to the partial $\kappa-\ln$ differentiation of the generalized ansatz (7) with respect to any component $v_i$, we obtain}
\begin{equation}\label{phichi}
\Phi(\chi)=\frac{1}{v_i}\frac{d}{dv_i}[\ln_\kappa f(v_i)]\;\;.
\end{equation}

The second member of Eq. (\ref{phichi}) depends only on $v_i$, with
$i=\emph{x, y or z}$.
Hence, Eq. (\ref{phieq}) can be satisfied only if all its members are
equal to one and the same constant, not depending on any of the velocity
components. Thus we can make $\Phi(\chi)=-m\gamma$, obtaining
\begin{equation}\label{diferential}
\frac{1}{v_i}\frac{d}{dv_i}[\ln_\kappa f(v_i)]=-m\gamma\;,
\end{equation}
where $\gamma=\frac{1}{k_BT}$ .\\

{\bf The solution of Eq. (\ref{diferential}) is given by}
\begin{equation}
\ln_\kappa f(v_i)=-\frac{m\gamma v_i^2}{2}\;\;,
\end{equation}
which provides
\begin{equation}\label{fvi0}
f(v_i)=\exp_\kappa \left(-\frac{m v^2_i}{2k_BT}\right)\;,
\end{equation}
In order to calculate the complete distribution, we insert the expressions (\ref{fvi0})
into Eq. (\ref{kdistribution}) to obtain
\begin{equation}\label{fkk}
F(v)=\left[\sqrt{1+\kappa^2\left(-\frac{m
v^2}{2 k_B T}\right)^2}-\frac{\kappa m v^2}{2k_B T}\right]^\frac{1}{\kappa}\;.
\end{equation}

{\bf If we assume the function $f(v_i)$ to be normalizable, we can write
\begin{equation}\label{fvi}
f(v_i)={1\over Z}\exp_\kappa \left(-\frac{m v^2_i}{2k_BT}\right)\;,
\end{equation}
where $1/Z$ is the $\kappa$-normalization constant.}

In order to calculate the so-called $\kappa$-normalization, let us introduce the expression \cite{k02}
\begin{equation}\label{nk00}
Z=\int_{\mathcal{R}} d^nv \exp_\kappa\left(-\frac{m v^2}{2k_BT}\right)\;,
\end{equation}
where $n$ (=1,2,3) is the number of the degree of freedom. Here, considering $n=1$, $a=\frac{m}{2k_BT}$ and $x=av^2$, we obtain
\begin{equation}
Z=\frac{1}{\sqrt{a}}\int_0^\infty x^{-1/2}\exp_\kappa(-x)dx.
\end{equation}
Using the generalized gamma functions \cite{k02}
\begin{eqnarray}\label{kgamma}\nonumber
\int_0^\infty x^{r-1}\exp_\kappa(-x)dx &=& \frac{[1+(r-2|\kappa|)|2\kappa|^r]}{[1-(r-1)|\kappa|]^2-\kappa^2}\\
&&\times\frac{\Gamma\left(\frac{1}{|2\kappa|}-\frac{r}{2}\right)}{\Gamma\left(\frac{1}{|2\kappa|}+\frac{r}{2}\right)}
\Gamma\left(\frac{1}{2}\right),
\end{eqnarray}
and after some algebra, we obtain
\begin{equation}
Z=\sqrt{\frac{\pi k_BT}{m}}\frac{|\kappa|^{-1/2}}{\left[1-{\frac{1}{2}}|\kappa|\right]}
\frac{\Gamma\left(\frac{1}{|2\kappa|}-\frac{1}{4}\right)}{\Gamma\left(\frac{1}{|2\kappa|}+\frac{1}{4}\right)}\;.
\end{equation}
It is easy to see that the standard Maxwellian result $(1/Z)=(m/2\pi k_BT)^{1/2}$ is recovered in the gaussian limit $\kappa=0$.

Using the expression (\ref{nk00}) for $n=3$, we can show that the normalization for the complete distribution is given by
\begin{equation}
\frac{1}{Z}=\left(\frac{m|\kappa|}{\pi k_BT}\right)^{3/2}\left[1+\frac{3}{2}|\kappa|\right]
\frac{\Gamma\left(\frac{1}{|2\kappa|}+\frac{3}{4}\right)}{\Gamma\left(\frac{1}{|2\kappa|}-\frac{3}{4}\right)}\;,
\end{equation}
As expected, the $\kappa$-distribution above is isotropic meaning that all velocity directions are equivalent in this generalized context. Here, we emphasize that in the limit $\kappa =0$ both the
normalization and the distribution function reproduce the standard Maxwellian (\ref{1}).

\section{Non-gaussian framework and stellar polytropes}

As mentioned in the introduction, we shall discuss a connection between the Kaniadakis statistics and polytropic index in
the context of the self-gravitating system. Following a procedure considered in Ref. \cite{plastino93,plastino05}, let us start with the Kaniadakis generalized entropy of index $\kappa$ of the form
\begin{equation}
 S_{\kappa}(f)=-\int \frac{f^{1+\kappa}-f^{1-\kappa}}{2\kappa}d\Omega,
\end{equation}
where $d\Omega=d^3\mbox{\boldmath $r$}d^3\mbox{\boldmath $v$}$, and the parameter $\kappa\neq 0$ provides a possible generalization of
the Boltzmann-Gibbs entropy. The extremum entropy state can be derived by varying $S_\kappa$ with respect to $f$. Using
the Lagrange multipliers $\alpha$ and $\beta$, the extremum solution subject to constraints \footnote{The mass and energy total of the self gravitating system governed by the Vlasov and Poisson equations. For more details, see Ref. \cite{binney87}} $M=\int f d\Omega$ and
$E=K+U={1\over 2}\int v^2  f d\Omega + {1\over 2}\int \phi f d\Omega$  is obtained from
\begin{equation}\label{vars}
\delta S_\kappa -\alpha \delta M - \beta \delta E = 0
\end{equation}
which leads to
\begin{eqnarray}\nonumber
\int \left\{-\frac{1}{2\kappa}\right.[(1&+&\kappa)f^{\kappa}-(1-\kappa)f^{-\kappa}]-\alpha \\
 &&\left.-\beta \left({v^2\over 2}
+\phi\right)\right\}\delta f d\Omega=0.
\end{eqnarray}

We have used the relation $\int \delta\phi f d\Omega= \int\phi\delta f d\Omega$ for derivation of above expression. Here, $\phi$ denotes gravitational potential, and the equation (\ref{vars}) must be satisfied independently of the choice of $\delta f$. Thus, we obtain the following distribution function
\begin{equation}\label{fk001}
f(\mbox{\boldmath $r$},\mbox{\boldmath $v$})=B \left[ \phi_0(\epsilon) -\phi (\mbox{\boldmath $r$})- {v^2\over 2}  \right]^{1/\kappa}
\end{equation}
where the constants $B$ and $\phi_0$ are defined by
\begin{equation}\nonumber
B=\left( {\kappa\beta\over (1+\kappa)} \right)^{1/\kappa}\;\;,
\end{equation}
\begin{equation}
\phi_0(\epsilon)=\frac{-{\sqrt{\kappa^2(-\alpha-\beta\epsilon)^2+1-\kappa^2}-\alpha\kappa}}{\beta\kappa},
\end{equation}
{\bf and $\epsilon={v^2\over 2} + \phi$.}

{\bf In order to obtain a relation between the stellar polytrope index $n$ and entropic parameter $\kappa$, let us now introduce the polytropic sphere distribution
\begin{equation}\label{eeee}
f\sim\mathcal{E}^{n-3/2},
\end{equation}
where $\mathcal{E}$ is the relative energy of a star, given by \cite{binney87}
\begin{equation}
\mathcal{E}=\Psi-{1\over 2} v^2,
\end{equation}
and $\Psi$ is the relative potential of a star associated with $\phi$. Therefore, comparing the distribution (\ref{fk001}) with (\ref{eeee}), we have}
\begin{equation}\label{nk}
 n=\frac{3}{2}+\frac{1}{\kappa}.
\end{equation}

{\bf Let us now compare the expression $n=n(\kappa)$ with the Tsallis relation $n=n(q)$. In this regard, Plastino and Plastino \cite{plastino93,plastino05} have introduced an expression given by}
\begin{equation}\label{qindex}
 n=\frac{3}{2}+\frac{1}{q-1},
\end{equation}
which includes the isothermal situation for $n
\rightarrow \infty$, i.e., the Gaussian limit $q=1$. In order to guarantee the conservation of mass and energy, the Tsallis' parameter $q$
should be larger than 9/7.

\begin{figure}[h]
\centerline{\includegraphics[angle=-90, width=12cm]{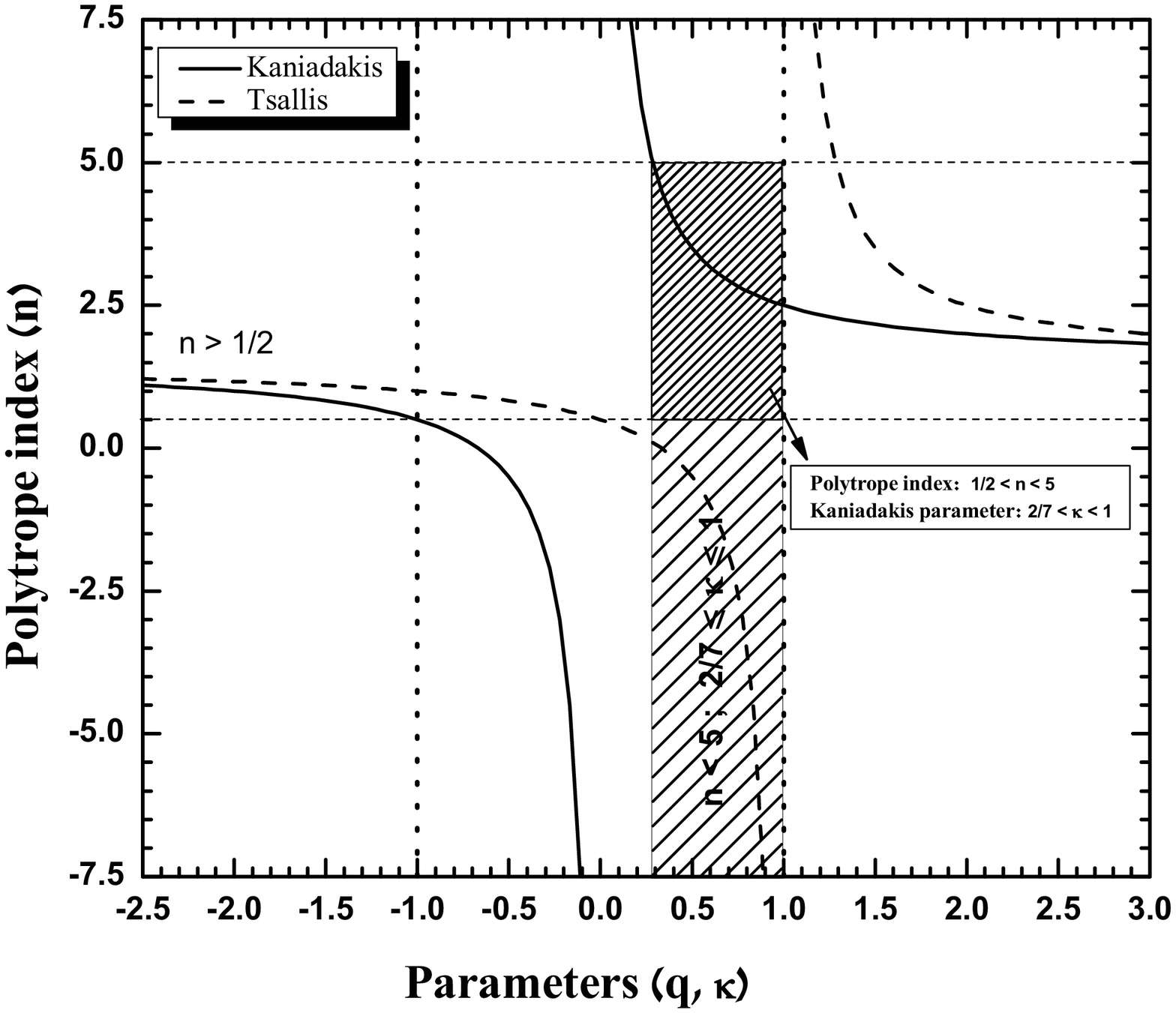}}
\caption{\footnotesize Polytrope index $n$ as a function of the Kaniadakis ($\kappa$) and Tsallis ($q$) parameters. The vertical dotted lines mark the limits of validity of $\kappa$-statistical ($-1\leq \kappa \leq 1$)\cite{k01,k02,k03,review,k10,k11}.}
\end{figure}

In Figure $1$ we display the dependence between the entropic indexes  ($\kappa;q$) and the polytropic index ($n$), given by Eqs. (\ref{nk}) and (\ref{qindex}), respectively. We show that the polytropic index diverges for the non-Gaussian parameters $\kappa= 0,q=1$, i.e., the isothermal spheres separate the polytropes into two branches. For $\kappa>0$ and $q>1$, $n$ ranges from $\infty$ to $3/2$. From $f\sim\mathcal{E}^{n-3/2}$, we see that $\mathcal{E}$ is a decreasing function of the energy and, when $n=3/2$, the distribution function becomes a constant independent of energy. On the other hand, it is well known that, for any astrophysical system, $n$ should be positive and higher than $1/2$ \cite{binney87}. Thus the intervals $0\leq q \leq 1$ and $-1\leq \kappa \leq 0$ represent forbidden regions, since the values of $n$ index tend to be smaller than $1/2$ and negative.

It is also worth observing that $\kappa<-1$ and $q<0$ also provides $n>1/2$. However this limits on $\kappa;q$ violates the validity of $\kappa$-statistics \cite{k01,k02,k03,review,k10,k11} and the non-extensive $H$-theorem \cite{TH}, respectively. As is well known, for the polytropic index  $n>5$, the density falls off so slowly at large radii that the mass is infinite \cite{binney87}. Figure 1 shows that only the branches on the right-side are physically significant for $n<5$. Therefore, the physical values of the Tsallis parameter $q$ should be larger than $9/7$\footnote{Investigating a collisionless stellar gas in the context of Non-extensive Kinetic Theory, the bound $q>9/7$ also was calculated in Ref. \cite{lima05}}, and the Kaniadakis parameter $\kappa$ should be constrained to interval of validity $\kappa\in[2/7;1]$ represented by the lower dashed rectangle. {\bf It is also worth noting that the polytropic indices obtained from Kaniadakis' statistics are restricted to the range $5/2\leq n\leq 5$, which excludes important stellar polytopes of index $n = 3/2$, i.e., the models of adiabatic stars supported by pressure of a non-relativistic gas \cite{chandra}.}

{\bf Summing up, a close examination of
Figure 1 tells us that the two distributions studied here
present a similar behaviour for stellar polytropes,
although Kaniadakis function is more restrictive than
Tsallis. In reality, in order to know that framework is better, we should do a study based on observational data, e.g., an investigation considering the comparison between Stellar polytropes and
Navarro-Frenk-White halo models for the description of dark matter halos (See Ref. \cite{navarro}). In this regard, an analysis considering this issue is currently under investigation.}

\section{Conclusions}

 In this work we have studied Kaniadakis statistics based on the generalized Maxwellian formulation for the $\kappa$-statistics and, as an application, we investigate the physical effect on the stellar polytropic system. In the first part, we conclude that there is a Kanidakis velocity distribution given by Eq. (\ref{fkk}) that is uniquely determined by the requirements of (i) isotropy and (ii) a generalization of the factorizability condition. From the physical viewpoint, this generalization introduces statistical {\bf dependence} between the velocity components, when we replace the logarithm function by a power law (a similar argument was considered in Refs. \cite{TH,TH1,TH2,TH3,santos11} for the non-Gaussian generalization of molecular chaos hypothesis). In particular, Maxwell expressions are recovered in the Gaussian limit, $\kappa\rightarrow 0$.

 From an application perspective, we have shown that the expressions for Tsallis and Kaniadakis, given by Eqs. (\ref{qindex}) and (\ref{nk}), present similar behaviours. However, the astrophysical limit on the polytropic index $1/2<n<5$, provides the constraint $\kappa\in[2/7;1]$ and $q>9/7$ for the Kaniadakis and Tsallis parameters, respectively. It is worth mentioning that the Gaussian limit $\kappa=0$, equivalent the Tsallis expression Eq. (\ref{qindex}) for $q=1$, reproduces Maxwellian isothermal spheres or, equivalently, $n =\infty$.

\vspace{0,5cm}
\begin{acknowledgments}
We would also like to thank the anonymous referees for valuable suggestions
and comments. The authors thank CNPq for the grants under which this work was
carried out. R.S. and J.R.P.S also thank financial support
from INCT-INEspa\c co.

\end{acknowledgments}

\end{document}